\PassOptionsToPackage{unicode}{hyperref}
\PassOptionsToPackage{hyphens}{url}
\PassOptionsToPackage{dvipsnames,svgnames,x11names}{xcolor}
\documentclass[
  12pt]{article}

\usepackage{amsmath,amsthm,amsbsy,amsfonts}
\usepackage{bm,graphicx,psfrag,epsf}
\usepackage{enumerate}
\usepackage[round, authoryear]{natbib}
\usepackage{url} 
\usepackage{bbm}
\usepackage{ifpdf}
\usepackage{xr-hyper}

\usepackage{setspace}
\usepackage[usenames]{color}
\RequirePackage[OT1]{fontenc}
\usepackage{algorithm}
\usepackage{algpseudocode}

\usepackage{mathtools}

\usepackage{booktabs}

\usepackage{paralist}

\usepackage{amsbsy}

\usepackage[mathscr]{eucal}

\usepackage{bm}

\usepackage{xspace}

\usepackage{multirow}

\usepackage{subcaption} 

\mathchardef\mhyphen="2D

\ifpdf
\usepackage[colorlinks,urlcolor=blue,citecolor=blue,linkcolor=blue]{hyperref}
\else
\newcommand{\href}[2]{{#2}}
\usepackage{url}
\fi

\ifpdf
\newcommand{\Sec}[1]{\hyperref[sec:#1]{Section~\ref*{sec:#1}}} 
\newcommand{\App}[1]{\hyperref[sec:#1]{Appendix~\ref*{sec:#1}}} 
\newcommand{\Supp}[1]{\hyperref[sec:#1]{Supplement~\ref*{sec:#1}}} 
\newcommand{\Eqn}[1]{\hyperref[eq:#1]{{\rm (\ref*{eq:#1})}}} 
\newcommand{\Part}[1]{\hyperref[part:#1]{(\ref*{part:#1})}} 
\newcommand{\Fig}[1]{\hyperref[fig:#1]{Figure~\ref*{fig:#1}}} 
\newcommand{\Tab}[1]{\hyperref[tab:#1]{Table~\ref*{tab:#1}}} 
\newcommand{\Thm}[1]{\hyperref[thm:#1]{Theorem~\ref*{thm:#1}}} 
\newcommand{\Lem}[1]{\hyperref[lem:#1]{Lemma~\ref*{lem:#1}}} 
\newcommand{\Prop}[1]{\hyperref[prop:#1]{Proposition~\ref*{prop:#1}}} 
\newcommand{\Cor}[1]{\hyperref[cor:#1]{Corollary~\ref*{cor:#1}}} 
\newcommand{\Def}[1]{\hyperref[def:#1]{Definition~\ref*{def:#1}}} 
\newcommand{\Alg}[1]{\hyperref[alg:#1]{Algorithm~\ref*{alg:#1}}} 
\newcommand{\Ex}[1]{\hyperref[ex:#1]{Example~\ref*{ex:#1}}} 
\newcommand{\As}[1]{\hyperref[as:#1]{Assumption~{\rm\ref*{as:#1}}}} 
\newcommand{\Reg}[1]{\hyperref[as:#1]{Condition~\ref*{reg:#1}}} 
\newcommand{\AlgLine}[2]{\hyperref[alg:#1]{line~\ref*{line:#2} of Algorithm~\ref*{alg:#1}}}
\newcommand{\AlgLines}[3]{\hyperref[alg:#1]{lines~\ref*{line:#2}--\ref*{line:#3} of Algorithm~\ref*{alg:#1}}}
\else
\newcommand{\Sec}[1]{{Section~\ref{sec:#1}}} 
\newcommand{\App}[1]{{Appendix~\ref{sec:#1}}} 
\newcommand{\Supp}[1]{{Supplement~\ref{sec:#1}}} 
\newcommand{\Eqn}[1]{{(\ref{eq:#1})}} 
\newcommand{\Part}[1]{{(\ref{part:#1})}} 
\newcommand{\Fig}[1]{{Figure~\ref{fig:#1}}} 
\newcommand{\Tab}[1]{{Table~\ref{tab:#1}}} 
\newcommand{\Thm}[1]{{Theorem~\ref{thm:#1}}} 
\newcommand{\Lem}[1]{{Lemma~\ref{lem:#1}}} 
\newcommand{\Prop}[1]{{Proposition~\ref{prop:#1}}} 
\newcommand{\Cor}[1]{{Corollary~\ref{cor:#1}}} 
\newcommand{\Def}[1]{{Definition~\ref{def:#1}}} 
\newcommand{\Alg}[1]{{Algorithm~\ref{alg:#1}}} 
\newcommand{\Ex}[1]{{Example~\ref{ex:#1}}} 
\newcommand{\Reg}[1]{{R~\ref*{reg:#1}}} 
\fi

\newcommand{\Real}{\mathbb{R}}

\newcommand{\Tra}{^{\sf T}} 

\newcommand{\V}[1]{{\bm{\mathbf{\MakeLowercase{#1}}}}} 

\newcommand{\M}[1]{{\bm{\mathbf{\MakeUppercase{#1}}}}} 

\definecolor{blue}{rgb}{0.2,0.5,0.7}
\definecolor{green}{rgb}{0.3,0.68,0.29}
\definecolor{purple}{rgb}{0.6,0.31,0.64}

\usepackage{iftex}
\ifPDFTeX
  \usepackage[T1]{fontenc}
  \usepackage[utf8]{inputenc}
  \usepackage{textcomp} 
\else 
  \usepackage{unicode-math}
  \defaultfontfeatures{Scale=MatchLowercase}
  \defaultfontfeatures[\rmfamily]{Ligatures=TeX,Scale=1}
\fi
\usepackage{lmodern}
\ifPDFTeX\else  
\fi
\IfFileExists{upquote.sty}{\usepackage{upquote}}{}
\IfFileExists{microtype.sty}{
  \usepackage[]{microtype}
  \UseMicrotypeSet[protrusion]{basicmath} 
}{}
\makeatletter
\@ifundefined{KOMAClassName}{
  \IfFileExists{parskip.sty}{%
    \usepackage{parskip}
  }{
    \setlength{\parindent}{0pt}
    \setlength{\parskip}{6pt plus 2pt minus 1pt}}
}{
  \KOMAoptions{parskip=half}}
\makeatother
\usepackage{xcolor}
\makeatletter
\ifx\paragraph\undefined\else
  \let\oldparagraph\paragraph
  \renewcommand{\paragraph}{
    \@ifstar
      \xxxParagraphStar
      \xxxParagraphNoStar
  }
  \newcommand{\xxxParagraphStar}[1]{\oldparagraph*{#1}\mbox{}}
  \newcommand{\xxxParagraphNoStar}[1]{\oldparagraph{#1}\mbox{}}
\fi
\ifx\subparagraph\undefined\else
  \let\oldsubparagraph\subparagraph
  \renewcommand{\subparagraph}{
    \@ifstar
      \xxxSubParagraphStar
      \xxxSubParagraphNoStar
  }
  \newcommand{\xxxSubParagraphStar}[1]{\oldsubparagraph*{#1}\mbox{}}
  \newcommand{\xxxSubParagraphNoStar}[1]{\oldsubparagraph{#1}\mbox{}}
\fi
\makeatother

\usepackage{longtable,booktabs,array}
\usepackage{calc} 
\usepackage{etoolbox}
\makeatletter
\patchcmd\longtable{\par}{\if@noskipsec\mbox{}\fi\par}{}{}
\makeatother
\IfFileExists{footnotehyper.sty}{\usepackage{footnotehyper}}{\usepackage{footnote}}
\makesavenoteenv{longtable}
\makeatletter
\def\maxwidth{\ifdim\Gin@nat@width>\linewidth\linewidth\else\Gin@nat@width\fi}
\def\maxheight{\ifdim\Gin@nat@height>\textheight\textheight\else\Gin@nat@height\fi}
\makeatother
\setkeys{Gin}{width=\maxwidth,height=\maxheight,keepaspectratio}
\makeatletter
\def\fps@figure{htbp}
\makeatother

\makeatletter
\@ifpackageloaded{caption}{}{\usepackage{caption}}
\AtBeginDocument{%
\ifdefined\contentsname
  \renewcommand*\contentsname{Table of contents}
\else
  \newcommand\contentsname{Table of contents}
\fi
\ifdefined\listfigurename
  \renewcommand*\listfigurename{List of Figures}
\else
  \newcommand\listfigurename{List of Figures}
\fi
\ifdefined\listtablename
  \renewcommand*\listtablename{List of Tables}
\else
  \newcommand\listtablename{List of Tables}
\fi
\ifdefined\figurename
  \renewcommand*\figurename{Figure}
\else
  \newcommand\figurename{Figure}
\fi
\ifdefined\tablename
  \renewcommand*\tablename{Table}
\else
  \newcommand\tablename{Table}
\fi
}
\@ifpackageloaded{float}{}{\usepackage{float}}
\floatstyle{ruled}
\@ifundefined{c@chapter}{\newfloat{codelisting}{h}{lop}}{\newfloat{codelisting}{h}{lop}[chapter]}
\floatname{codelisting}{Listing}

\makeatother
\makeatletter
\@ifpackageloaded{caption}{}{\usepackage{caption}}
\@ifpackageloaded{subcaption}{}{\usepackage{subcaption}}
\makeatother

\ifLuaTeX
  \usepackage{selnolig}  
\fi
\usepackage{bookmark}

\IfFileExists{xurl.sty}{\usepackage{xurl}}{} 
\hypersetup{
  pdftitle={Title},
  pdfauthor={Author 1; Author 2},
  pdfkeywords={3 to 6 keywords, that do not appear in the title},
  colorlinks=true,
  linkcolor={blue},
  filecolor={Maroon},
  citecolor={Blue},
  urlcolor={Blue},
  pdfcreator={LaTeX via pandoc}}

\newcommand{\anon}{1}

 \makeatletter
\newcommand*{\addFileDependency}[1]{
  \typeout{(#1)}
  \@addtofilelist{#1}
  \IfFileExists{#1}{}{\typeout{No file #1.}}
}
\makeatother

\begin{document}

\def\spacingset#1{\renewcommand{\baselinestretch}%
{#1}\small\normalsize} \spacingset{1}


\if1\anon
{
  \title{\bf Transfer Learning for Robust Structured Regression with Bi-level Source Detection}
\author{
   Haoming Shi \\
    Department of Statistics,
    Rice University\\
    Yang Feng\\
    Department of Biostatistics,  New York University\\
     Xiaoqian Liu\\
    Department of Statistics,
    University of California at Riverside }
\date{}
  \maketitle
} \fi

\if0\anon
{
  \bigskip
  \bigskip
  \bigskip
  \begin{center}
    {\LARGE\bf Transfer Learning for Robust Structured Regression with Bi-level Source Detection}
\end{center}
  \date{}
  \medskip
} \fi

\bigskip
\begin{abstract}
High-dimensional data in modern applications, such as COVID-19 mortality, often span multiple domains. Leveraging auxiliary information from source domains to improve performance in a target domain motivates the use of transfer learning.
However, a practical issue that has
been overlooked is data contamination, which induces heterogeneity and can significantly degrade transfer learning performance. 
To address this challenge, we propose a novel approach that tackles transfer learning under data contamination within a structured regression setting.  By employing the robust L\textsubscript{2}E  criterion, we develop the TransL\textsubscript{2}E method that accounts for contamination in both target and source data while effectively transferring relevant information. Beyond robust estimation,  TransL\textsubscript{2}E introduces a data-driven bi-level source detection mechanism, operating at both individual and cohort levels, which possesses multiple advantages over existing source detection approaches.  
Comprehensive simulation studies and a real data application demonstrate the superior performance of TransL\textsubscript{2}E in both robust estimation and structure recovery in the presence of data limitation and contamination.
\end{abstract}

\noindent%
\vfill

\newpage
\spacingset{1.8} 

\section{Introduction}
\label{sec: intro}

Modern data applications often involve observations collected across related yet heterogeneous domains, where leveraging auxiliary information can substantially enhance estimation and prediction in a target population, particularly when target data are limited. A representative example, as studied in Section \ref{sec:real-data}, arises in modeling COVID-19 mortality using county-level data across the United States. Due to shared demographic, socioeconomic, and healthcare characteristics, neighboring regions often exhibit similar patterns, suggesting that borrowing information across regions can improve statistical efficiency. Such settings naturally motivate transfer learning approaches \citep{pan2009survey}, which exploit cross-domain similarities to improve learning performance in the target domain.

However, real-world data rarely satisfy the idealized assumptions required for effective transfer. In practice, outliers, and more generally data contamination, are prevalent across domains,  which introduce heterogeneity and result in substantial deviations from shared patterns. This creates a fundamental challenge: naively aggregating information may amplify heterogeneity and degrade performance in the target domain (i.e., negative transfer \citep{torrey2010transfer}), while discarding entire data sources risks losing valuable information that could otherwise improve estimation. Consequently, a key question is how to selectively leverage useful information from heterogeneous and contaminated sources to achieve robust improvements in the target task. Addressing this challenge requires methods that can adaptively identify transferable information at a finer granularity, balancing robustness against contamination with efficient use of available data.

\subsection{Related work}

Transfer learning has emerged as a powerful paradigm for leveraging information from related source tasks to improve performance on a target task with limited data. 
Originating in computer science, 
transfer learning has recently attracted growing attention from the statistics community and been investigated in various statistical scenarios. For instance, \citet{bastani2021predicting} applied transfer learning to high-dimensional generalized linear models (GLMs) and proposed a two-step estimator to  improve prediction performance on a target task with limited data by leveraging a single source with abundant proxy data.  \citet{li2022transfer} studied transfer learning for high-dimensional linear regression with multiple source cohorts, establishing minimax-optimal error bounds for the proposed estimator and developing a data-driven procedure for informative source selection. 
\citet{tian2023transfer} further investigated a multi-source transfer learning
framework under the high-dimensional GLM settings, addressing both point estimation and confidence interval construction with theoretical guarantees. They also proposed an algorithm-free source detection approach to avoid negative transfer. In addition to high-dimensional (generalized) linear models, transfer learning has also found successful applications to a broad range of statistical models,  including Gaussian graphical models \citep{li2023transfer}, large-scale 
quantile regression \citep{jin2024transfer, li2024transfer},
Cox regression \citep{li2023accommodating, lu2025adaptive, liu2025transfer},  semiparametric regression \citep{hu2023optimal, he2024representation}, and nonparametric regression \citep{cai2024transfer}, among others.

Data contamination presents a critical yet understudied challenge in modern statistical learning. Contaminated observations, particularly outliers arising from measurement errors or mislabeling, can severely degrade the performance of statistical models. For example, least-squares regression
is known for its sensitivity to outliers, 
where even a single outlier can dramatically compromise its estimation accuracy \citep{zuo2023robust}.  Within the traditional structured regression paradigm (i.e., without incorporating any source data), a substantial body of work has focused on developing robust structured regression methods to handle contaminated data. To name a few, \cite{robust-isotonic} proposed a family of isotonic M-estimators to address corrupted observations in isotonic regression.  \cite{robust-convex} employed the absolute error loss for robust estimation in convex regression. Several studies \citep{She2011, MD-lasso, sparse-trim, Extended-Lasso} dealt with outliers in sparse linear regression by adopting different loss functions.  More recently, \cite{L2E-Chi} and \cite{L2E-Liu} developed a general framework for robust structured regression employing the $\text{L}_2\text{E}$ criterion \citep{L2E-Scott}, which automatically identifies outliers and mitigates their effects during the estimation process.  

\subsection{Our contributions}
This work addresses the aforementioned challenge of transfer learning under data contamination. We particularly  focus on structured linear regression as the target task and aim to simultaneously tackle data limitation and contamination. Specifically, we propose a novel method, named TransL\textsubscript{2}E, that integrates transfer learning with the L\textsubscript{2}E criterion for robust structured regression.   
TransL\textsubscript{2}E makes the following contributions to the literature:
\begin{enumerate}
    \item TransL\textsubscript{2}E applies the L\textsubscript{2}E criterion to both target and source datasets to account for contaminated observations, enabling robust coefficient estimation and structure recovery in transfer learning.
    
    \item By utilizing the case weights, a by-product of the L\textsubscript{2}E estimator,  TransL\textsubscript{2}E introduces a bi-level source detection mechanism that selects high-quality data at both individual and cohort levels, ensuring useful knowledge transfer and avoiding negative transfer.
    
    \item Beyond sparse regression with transfer learning,  TransL\textsubscript{2}E operates as a general and flexible structured regression framework, accommodating a broad range of structural assumptions. 
\end{enumerate}

\subsection{Organization}
The remainder of this article is organized as follows. In \Sec{background}, we review the robust L\textsubscript{2}E criterion and the structured L\textsubscript{2}E regression framework. In \Sec{method}, we set up the transfer learning problem and introduce the TransL\textsubscript{2}E method, including a detailed discussion of the bi-level source detection procedure.
In \Sec{simulation}, we present comprehensive simulation studies to evaluate the empirical performance of the  TransL\textsubscript{2}E method. In \Sec{real-data},  we apply TransL\textsubscript{2}E to a COVID-19 mortality prediction problem. We conclude with a discussion in \Sec{discussion}.

\section{Background}
\label{sec:background}

\subsection{The L\textsubscript{2}E criterion}
\label{sec:L2E}

We start with reviewing the L\textsubscript{2}E criterion under the parametric setting for robust estimation \citep{L2E-Scott}.  Let $\phi(y)$ be the unknown density function that we aim to estimate.  $\phi(y \mid \V \theta)$ is a probability density function parameterized by $\V \theta \in \Theta \subset \Real^p$ that approximates $\phi(y)$. The L\textsubscript{2}E criterion seeks an estimate of $\V \theta$ by minimizing the integrated squared error (ISE) or the $\text{L}_2$ distance between $\phi(y)$ and $\phi(y \mid \V \theta)$. Mathematically, the L\textsubscript{2} estimate (L\textsubscript{2}E) of $\V \theta$ solves 
\begin{equation}
\label{eq:ISE}
    \min_{\V \theta} \int \left[ \phi(y \mid \V \theta) - \phi(y)\right]^2 \,d y. 
\end{equation}
As $\phi(y)$ is unknown, it is practically intractable to recover $\V \theta$ by directly minimizing the ISE. Alternatively, we can minimize an unbiased estimate of this ISE in lieu of the exact one. To reveal this, we expand the quadratic integrand  in \eqref{eq:ISE} and rewrite the problem as 
\begin{equation}
\label{eq:ISE2}
     \min_{\V \theta} \int \phi(y \mid \V \theta)^2 \,dy - 2 \int \phi(y \mid \V \theta) \phi(y) \,dy + \int \phi( y)^2 \,d y. 
\end{equation}
Notice that the third integral in \eqref{eq:ISE2} is independent of $\V \theta$ and thus can be excluded from the minimization problem. The second integral is the expectation $E_Y[\phi(Y \mid \V \theta)]$, where $Y$ denotes a random variable drawn from $\phi$.  This quantity can be approximated by an unbiased estimate, e.g., its sample mean. The first integral finds a closed-form expression for many parametric models. All together, an approximate  L\textsubscript{2}E of $\V \theta$  is obtained by minimizing the following fully data-based L\textsubscript{2}E loss
\begin{eqnarray*}
    h(\V \theta) =  
    \int \phi(y \mid \V \theta)^2 \, d y - \frac{2}{n} \sum_{i=1}^n \phi(y_i\mid\V \theta),
\end{eqnarray*}
where $n$ denotes the sample size. L\textsubscript{2}E is a special case of a family of minimum-divergence estimators \citep{basu1998robust} which also includes the maximum likelihood estimator (MLE) as a limiting case.  MLE is the most efficient but the least robust;  L\textsubscript{2}E instead finds a reasonable balance between efficiency and robustness \citep{warwick2005}. 

Under the normal distribution assumption, $\phi(y \mid \V \theta)$  denotes the normal density function  with $\V \theta = (\mu, \tau)$, where $\mu$ is the mean parameter and $\tau$ is the precision parameter (i.e., the inverse of the standard deviation).  Explicitly, $\phi(y\mid \V \theta) = \phi(y\mid \mu, \tau) = \frac{\tau}{\sqrt{2\pi}}e^{-\frac{\tau^2(y-\mu)^2}{2}}$. It is easy to verify that $\int \phi(y \mid \V \theta)^2 \, d y = \frac{\tau}{2\sqrt{\pi}}$. Consequently, the  L\textsubscript{2}E of $\V \theta = (\mu, \tau)$ solves
\begin{eqnarray}
  \label{eq:L2E-normal}
 \min_{\V \theta = (\mu, \tau)} ~ h(\mu, \tau),
\end{eqnarray}
where $h(\mu, \tau)$ is expressed as
\begin{eqnarray*}
    h(\mu, \tau) = \frac{\tau}{2\sqrt{\pi}} - \frac{\tau}{n} \sqrt{\frac{2}{\pi}} \sum_{i=1}^{n} e^{-\frac{\tau^{2}(y_i-\mu)^{2}}{2}}.
\end{eqnarray*}
Since normal errors are commonly assumed in many statistical learning settings, this L\textsubscript{2}E \eqref{eq:L2E-normal} can serve as a robust alternative for MLE which corresponds to a least squares (LS) estimator. Notably,  L\textsubscript{2}E can achieve robustness even when the assumed parametric form is known to be incorrect \citep{L2E-Scott}. Therefore, in principle, any LS estimator can be robustified by the L\textsubscript{2}E \eqref{eq:L2E-normal}. Examples include structured linear regression \citep{L2E-Chi, L2E-Liu} and low-rank tensor decomposition \citep{heng2023robust}.

\subsection{Structured L\textsubscript{2}E regression}

Given a data set $\{y_i, \V x_i\}_{i=1}^n$, where $y_i$ is the $i$-th observed response and $\V x_i = (x_{i1}, \cdots, x_{ip})\Tra \in \Real^p$ is the $i$-th observed vector of covariates.  The classical linear regression model assumes
\begin{equation*}
    y_i = \V x_i\Tra \V \beta +  \tau^{-1}\epsilon_i, ~~~i=1, \cdots, n,
\end{equation*}
where $\V \beta \in \Real^p$ is an unknown vector of regression coefficients, $\tau \in \Real_+$ is an unknown precision parameter, and $\epsilon_i$s are i.i.d. standard Gaussian noises. By denoting $\V \theta = (\V \beta\Tra, \tau)\Tra$ we have $y_i \mid \V \theta \overset{\mathrm{i.i.d.}}{\sim}  N(\V x_i\Tra \V \beta, \tau^{-2})$, namely,
$$
\phi(y_i\mid \V \theta) =  \frac{\tau}{\sqrt{2\pi}}e^{-\frac{\tau^2r_i^2}{2}},
$$
where $r_i=y_i-\V x_i \Tra \V \beta$. 
According to \eqref{eq:L2E-normal}, the L\textsubscript{2}E of $\V \theta = (\V \beta\Tra, \tau)\Tra$ is given by 
\begin{eqnarray*}
  \label{eq:L2E-linear}
    \hat{\V \theta}_{\text{L}_2\text{E}}  =  \operatorname*{arg\,min}_{\V \theta = (\V \beta\Tra, \tau)\Tra} h(\V \beta, 
    \tau) = \operatorname*{arg\,min}_{\V \beta\in \Real^p, \tau \in \Real_+} ~ \frac{\tau}{2 \sqrt{\pi}}-\frac{\tau}{n} \sqrt{\frac{2}{\pi}} \sum_{i=1}^{n} e^{-\frac{\tau^{2}r_{i}^{2}}{2}}.
\end{eqnarray*}

Structured L\textsubscript{2}E regression \citep{L2E-Chi, L2E-Liu} introduces a set constraint on the vector of regression coefficients $\V \beta$ and solves the following constrained optimization problem
\begin{eqnarray}
\label{eq:L2E-constr}
    \min_{\V \beta \in \Real^p, \tau\in \Real_+} h(\V \beta, \tau), ~~~~\text{subject to}~~
    \bm \beta \in C,
\end{eqnarray}
where $C$ is the set constraint that defines the structure of $\V \beta$. For example, $C=\{\V \beta \in \Real^p : \lVert \V \beta \lVert_0 \leq k\}$ with some positive integer $k$ imposing sparsity on $\V\beta$. 
Equivalently, we can convert \eqref{eq:L2E-constr} into an unconstrained  problem 
\begin{eqnarray}
\label{eq:L2E-penalty}
      \min_{\V \beta \in \Real^p, \tau\in \Real_+} h(\V \beta, \tau) +
      \mathrm{P}(\V \beta),
\end{eqnarray}
where $\mathrm{P}$ is either the 0/$\infty$ indicator of the constraint set $C$ or a non-smooth penalty for violating the constraint. We refer readers to \cite{L2E-Chi} and \cite{L2E-Liu} for examples of  $\mathrm{P}$ that induce a wide range of structural assumptions on $\V\beta$.

Unlike the convex LS loss, the L\textsubscript{2}E  loss $h(\V \beta, \tau)$ in \eqref{eq:L2E-penalty} is non-convex and requires careful computational design. \cite{L2E-Liu} proposed an efficient computational framework for solving \eqref{eq:L2E-penalty}, which iteratively solves a penalized weighted LS problem. More importantly, \cite{L2E-Liu} 
provided an algorithmic interpretation of how
L\textsubscript{2}E robustifies the LS estimator through iteratively updating the weight associated with each observation.  As a by-product, the converged weights conveniently flag outliers, enabling automatic outlier detection within the L\textsubscript{2}E framework. 

\section{Method}
\label{sec:method}

\subsection{Problem setup}
\label{sec:set-up}

Suppose we have a target dataset $\{\V y^{(0)}, \M X^{(0)}\}$, where $\V y^{(0)} \in \Real^{n_0}$ is the vector of $n_0$ observed responses, and $\M X^{(0)} \in \Real^{{n_0}\times p}$ is the design matrix composed of $n_0$ observed vector of covariates $\V x_i^{(0)} \in \Real^p$ for $i=1, \cdots, n_0$. We consider a regression framework where the vector of regression coefficients is assumed to exhibit particular structures, such as sparsity \citep{tibshirani1996regression}, isotonicity \citep{barlow1972isotonic}, and group sparsity \citep{yuan2006model} , to name a few. 
We are interested in the challenging scenario where the sample size $n_0$ is small and the data are contaminated by outliers.  Our primary goal is to achieve robust coefficient estimation and structure recovery on the target dataset.

In addition to the limited and contaminated target data, we have access to $K$ related source datasets. The $k$-th source dataset is denoted by 
$\{\V y^{(k)}, \M X^{(k)}\}$, with response vector $\V y^{(k)} \in \Real^{n_k}$ and design matrix $\M X^{(k)} \in \Real^{{n_k}\times p}$, for $k=1, \cdots K$.  Typically, the source sample sizes $n_k$ are larger than $n_0$, but these datasets are also subject to contamination.

The goal here is to transfer useful information from the adequate but contaminated source data to boost the learning performance on the target task with limited and contaminated data. 
Achieving this objective requires addressing two key questions:
\begin{itemize}
    \item[(i)] What information can be transferred from the source data? In other words, what similarities exist between the target and source datasets?
    \item[(ii)] How to effectively transfer the information while accounting for contamination in both target and source datasets?
\end{itemize}

\subsection{The TransL\textsubscript{2}E method}
\label{sec:TransL2E}

We now introduce the transfer learning L\textsubscript{2}E (TransL\textsubscript{2}E) method to achieve robust structured regression on the target dataset by addressing the two aforementioned  questions. 

Regarding the first question, following the current literature, we assume similarities lie in the regression coefficients. Specifically, we assume the following linear regression model across target and source datasets 
\begin{equation*}
    y_i^{(k)} = {\V x_i^{(k)}}\Tra \V \beta^{(k)} + {\tau^{(k)}}^{-1} \epsilon_i, ~~~i=1, \cdots, n_k,
\end{equation*}
for $k=0, 1, \cdots K$. For the target coefficient vector $\V \beta^{(0)}$, it possess some special structure.  When the source coefficient vector $\V \beta^{(k)}$ is similar to $\V \beta^{(0)}$ in the sense that their difference $\V \beta^{(0)} - \V \beta^{(k)}$ is sparse (i.e., $\lVert \V \beta^{(0)} - \V \beta^{(k)}\rVert_q$ is reasonably small for $q \in [0,1]$),  the information of $\V \beta^{(k)}$ can be borrowed to enhance the learning of $\V \beta^{(0)}$ \citep{li2022transfer}. In this case,
the $k$-th source dataset is said to be transferrable or informative \citep{tian2023transfer}. 
Of note,  we do not assume each source coefficient vector $\V \beta^{(k)}$ shares exactly the same structure as $\V \beta^{(0)}$. Following \cite{he2024transfusion}, we refer to the setting where $\V \beta^{(0)}$ and $\V \beta^{(k)}$ differ as \textit{model shift}. 
In addition, we allow for differences between $\tau^{(0)}$ and $\tau^{(k)}$, which we refer to as \textit{precision shift}. Later in \Sec{simulation}, our simulation studies demonstrate the outperformance of  TransL\textsubscript{2}E under varying levels of model shift and precision shift. 

Addressing the second question sets the main bottleneck of the problem, since both target and source datasets are plagued by contamination. 
In the current literature where data are assumed to be clean,  transfer learning is usually done in two steps: (i) pool all target and source datasets together to obtain a preliminary estimator $\hat{\V u}$; and (ii) correct the bias  by estimating a sparse difference vector $\V \delta = \V \beta^{(0)}- \hat{\V u}$ using the target data and then compute the final estimate $ \hat{\V \beta}^{(0)} = \hat{\V \delta} + \hat{\V u}$. When data contamination presents in both target and source datasets,  pooling all data together could amplify the impact of contamination, resulting in an inaccurate estimate $\hat{\V u}$ even if using robust estimation methods such as  L\textsubscript{2}E, which will be illustrated in our simulation studies in \Sec{simulation}. Another challenge for transfer learning is the so-called transferrable source detection. In practice, it is not known a priori whether a source dataset is transferrable. Existing studies often split the target dataset and apply cross-validation-type procedures to identify transferable source datasets, see for instance \cite{tian2023transfer} and \cite{jin2024transfer}. However, such data-splitting strategies raise serious concerns regarding their practical applicability, especially when the target sample size is already limited. The issue becomes further exacerbated
in the presence of data contamination, where splitting a small and corrupted dataset may lead to unreliable source selection and degraded learning performance.

\begin{algorithm}
\caption{TransL\textsubscript{2}E}
\label{alg:TransL2E}
\textbf{Input:} Target dataset $\{\V y^{(0)}, \M X^{(0)}\}$, source datasets $\{\V y^{(k)}, \M X^{(k)}\}_{k=1}^K$.\\
\textbf{Output:}  The TransL\textsubscript{2}E estimate $\hat{\V \beta}_{\text{TransL\textsubscript{2}E}}^{(0)}$. 
 \begin{algorithmic}[1]
 
 \State \textbf{Step 1 (Bi-level source detection)}: \For{$k=1, \cdots, K$}
 
1) Merge the $k$-th source dataset with the target dataset into 
  
 $\V y^{(k+)} = \begin{pmatrix}
     \V y^{(0)}\\
     \V y^{(k)}
 \end{pmatrix}\in \Real^{(n_0+n_k)}$
 and
 $\M X^{(k+)} = \begin{pmatrix}
     \M X^{(0)}\\
     \M X^{(k)}
 \end{pmatrix}\in \Real^{(n_0+n_k)\times p}$.
 
2) Run the structured L\textsubscript{2}E regression model
 \begin{equation}
     \label{eq:merge}
     (\hat{\V \beta}_k, \hat{\tau}_k)= \operatorname*{arg\,min}_{\V \beta \in \Real^{p}, \tau\in \Real^+}\left\{h(\V \beta, \tau \mid \V y^{(k+)}, \M X^{(k+)}) + \lambda_k P(\V \beta)\right\}
 \end{equation}
where $\lambda_k>0$ is a regularization parameter chosen by cross-validation.  

3) Compute the weights $\V w^{(0)}$ and $\V w^{(k)}$ using the estimates $(\hat{\V \beta}_k, \hat{\tau}_k)$
\begin{align}
    w^{(0)}_i & = e^{-\hat{\tau}_k^2(y_i-{\V x_i^{(0)}}\Tra \hat{\V \beta}_k )^2/2}, ~~i=1, \cdots, n_0, \label{eq:w-t}\\
   w^{(k)}_j &= e^{-\hat{\tau}_k^2(y_j-{\V x_j^{(k)}}\Tra \hat{\V \beta}_k )^2/2}, ~~j=1, \cdots, n_k, \label{eq:w-s}
\end{align}
where $w^{(0)}_i$ is the weight associated with the $i$-th case in the target, and $w^{(k)}_j$ is the weight associated with the $j$-th case in the $k$-th source.

4) Conduct bi-level source detection as summarized in \Alg{TransL2E-selection} to get the selected source sample $\{\V y^{(k-)}, \M X^{(k-)}\}$. 
\EndFor

\State \textbf{Step 2 (Co-learning):} Merge all the selected source datasets $\{\V y^{(k-)}, \M X^{(k-)}\}_{k=1}^K$ with the complete target dataset $\{\V y^{(0)}, \M X^{(0)}\}$ to compute 
 \begin{equation}
     \label{eq:cotrain}
     (\hat{\V \beta}, \hat{\tau})= \operatorname*{arg\,min}_{\V \beta \in \Real^p, \tau\in \Real^+}\left\{ h(\V \beta, \tau \mid \V y^{(0)}, \M X^{(0)}, \V y^{(1-)}, \M X^{(1-)}, \cdots, \V y^{(K-)}, \M X^{(K-)}) + \lambda P(\V \beta)\right\}
 \end{equation}
where the regularization parameter $\lambda>0$ is selected by cross-validation. 

\State  \textbf{Step 3 (De-biasing):} Correct the potential bias in $\hat{\V \beta}$ by computing
  \begin{equation*}
      (\hat{\V \delta}, \hat{\tau} )= \operatorname*{arg\,min}_{\V \delta \in \Real^{p}, \tau \in \Real^+}\left\{ h(\V \beta, \tau \mid \V y^{(0)}-\M X^{(0)}\hat{\V \beta}, \M X^{(0)}) + \mu\lVert \V \delta\rVert_1\right\}
  \end{equation*}
 and obtain the final estimate $\hat{\V \beta}_{\text{TransL\textsubscript{2}E}}^{(0)} = \hat{\V \beta} +  \hat{\V \delta}$, where the regularization parameter $\mu>0$ is selected by cross-validation.
 
\end{algorithmic} 
\end{algorithm}

Our TransL\textsubscript{2}E method distinguishes itself in effectively addressing these challenges through a principled strategy that selects high-quality source data at both the individual and cohort levels.
Specifically, the TransL\textsubscript{2}E procedure can be outlined in the following three steps. First, for each $k=1, \cdots, K$, we merge the $k$-th source dataset with the target dataset to obtain $(\hat{\V \beta}_k, \hat{\tau}_k)$ by running the structured L\textsubscript{2}E regression model \eqref{eq:merge}. The converged weights computed using $(\hat{\V \beta}_k, \hat{\tau}_k)$  will be utilized to conduct bi-level source detection, 
which we discuss later in Section \ref{sec:bi-level}. Second, we pool all the selected source data and the target data to co-learn an estimate $\hat{\V \beta}$ from model \eqref{eq:cotrain}. At last, we correct the potential bias in $\hat{\V \beta}$ by fitting a sparse L\textsubscript{2}E regression model on $\V \delta = \V \beta^{(0)} - \hat{\V \beta}$ using the target data. The complete TransL\textsubscript{2}E framework is presented in \Alg{TransL2E}.

We highlight some important details in our designed TransL\textsubscript{2}E framework. First, in the bi-level source detection step, we conduct the selection on each source dataset separately using the complete target dataset as a reference. We deliberately avoid pooling all source datasets together to mitigate data heterogeneity  and prevent amplifying the contamination effects in the source data. 
 More importantly, our approach does not require splitting the target dataset to achieve source detection, thereby providing more reliable selection compared to cross-validation-type approaches.  
Second, we incorporate the whole  target dataset in the co-learning step to make full use of the limited data and guarantee its maximum contribution. 
The contamination issue in the target dataset will be handled by the L\textsubscript{2}E criterion. 
Third, although  L\textsubscript{2}E jointly estimates both the coefficient parameter $\V \beta$ and the precision parameter $\tau$, our main focus lies in accurately estimating $\V \beta$. During the bi-level source detection step,  
the precision estimates $\hat{\tau}_k$s are required for computing the converged weights, whereas in the remainder of the procedure, $\tau$ is essentially treated as a nuisance parameter.


\subsection{Bi-level source detection}
\label{sec:bi-level}

We now describe how we utilize the converged case weights $\V w^{(0)}$ and $\V w^{(k)}$, computed from \eqref{eq:w-t} and \eqref{eq:w-s}, to perform bi-level source detection within the TransL\textsubscript{2}E framework. 

We begin by revisiting  the interpretation of the converged weights in the L\textsubscript{2}E regression \citep{L2E-Liu}. Suppose $(\hat{\V \beta}, \hat{\tau})$ denotes the L\textsubscript{2}E estimates obtained from the data $\{\V y \in \Real^n, \M X \in \Real^{n \times p}\}$. The $i$-th case $(y_i, \V x_i)$ has its associated weight defined by $w_i = e^{-\hat{\tau}^2(y_i-{\V x_i}\Tra \hat{\V \beta} )^2/2}$ for $i=1, \cdots, n$. A small weight $w_i$ implies the $i$-th case makes little contribution to the L\textsubscript{2}E regression. In other words, L\textsubscript{2}E identifies it as a potential outlier.

In the bi-level source detection step as shown in \Alg{TransL2E}, we merge the target dataset $\{\V y^{(0)}, \M X^{(0)}\}$ and the $k$-th source dataset $\{\V y^{(k)}, \M X^{(k)}\}$ to get the combined dataset $\{\V y^{(k+)}, \M X^{(k+)}\}$. If the true vectors of coefficients $\V \beta^{(0)}$ and $\V \beta^{(k)}$ are sufficiently close, fitting the L\textsubscript{2}E model \eqref{eq:merge} produces an estimate  $\hat{\V \beta}_{k}$ that approximates both $\V \beta^{(0)}$ and $\V \beta^{(k)}$. The resulting weight vector $\V w^{(k)}$ then flags outliers in 
the $k$-th source dataset. To address the data contamination issue, a straightforward approach is to apply a hard threshold to $\V w^{(k)}$ to select high-quality data points from the $k$-th source dataset. In this work, however, we instead leverage both  $\V w^{(0)}$ and $\V w^{(k)}$  in a probabilistic manner to perform data selection at both individual and cohort levels. 



By definition, each weight reflects how closely the corresponding case aligns with the fitted model \eqref{eq:merge}. As such, the empirical density function of the weights can be viewed as a proxy for the underlying data quality distribution. Building on this idea, we propose to sample data points from the $k$-th source dataset $\{\V y^{(k)}, \M X^{(k)}\}$ via a pseudo-importance sampling procedure. We treat the estimated  density of the source weights, denoted by $\hat{f}_k$, as our ``proposal'' distribution, and the estimated  density of the target weights, $\hat{f}_0$, as the ``target'' distribution from which we aim to sample.

The procedure operates as follows. Given the target weights $\{w_i^{(0)}\}_{i=1}^{n_0}$ and source weights $\{w_j^{(k)}\}_{j=1}^{n_k}$, we first estimate $\hat{f}_0$ and $\hat{f}_k$ using the kernel density estimation method \citep{scott2015multivariate} and calculate their Hellinger distance $h_k$ as a global measure of cohort-level divergence. Next, we construct the pseudo-importance weight $p_j$ for each source sample case $j$ as 
\begin{equation}
\label{eq:sample-weight}
    p_j = w_j^{(k)} \times 
 \frac{\hat{f}_0(w^{(k)}_j)}{\hat{f}_k(w^{(k)}_j) + \epsilon} \times \exp{(-\sqrt{h_k})}, ~~ j=1, \cdots, n_k
\end{equation}
where $\epsilon > 0$ is a small constant to avoid division by $0$. In our implementation, we set $\epsilon = 10^{-8}$, and the selection results are robust to the choice of $\epsilon$ provided it remains sufficiently small. Conceptually, the weight $w_j^{(k)}$ serves as the baseline importance weight of each source data point. 
The ratio $\frac{\hat{f}_0(w^{(k)}_j)}{\hat{f}_k(w^{(k)}_j) + \epsilon}$ adjusts this importance by referencing the target distribution. It upweighs those source data points that are common in the target but underrepresented in the source, and downweighs points common in the source but rare in the target. The term $\exp{(-\sqrt{h_k})}$
further modulates the importance weight based on the cohort-level distributional similarity.
Finally, a classical acceptance-rejection procedure is performed by comparing each $p_j$ to a uniform draw $u_j \sim \text{Unif}(0,1)$. \Alg{TransL2E-selection} presents the complete sampling procedure.


\begin{algorithm}[t]
\caption{Source selection for TransL\textsubscript{2}E}
\label{alg:TransL2E-selection}
\textbf{Input:} Target weight vector $\V w^{(0)} \in \Real^{n_0}$, the $k$-th source dataset $\{\V y^{(k)}, \M X^{(k)}\}$ and its associated  weight vector $\V w^{(k)} \in \Real^{n_k}$. \\
\textbf{Output:}  Selected source sample  $\{\V y^{(k-)}, \M X^{(k-)}\}$.
\begin{algorithmic}[1]    
    
\State Estimate the density functions of the target and source weights, denoted by $\hat{f}_0$ and $\hat{f}_k$, using the weight vectors $\V w^{(0)}$ and $\V w^{(k)}$, respectively.
    
    \State Compute the Hellinger distance between $\hat{f}_0$ and $\hat{f}_k$, denoted by $h_k$.
    \For{$j=1, \cdots, n_k$}
        \State Compute        
        $$p_j = \, w_j^{(k)} \times \frac{\hat{f}_0(w^{(k)}_j)}{\hat{f}_k(w^{(k)}_j) + \epsilon} \times  \exp{(-\sqrt{h_k})} ,$$        
        where $\epsilon>0$ is a small constant to avoid division by 0.
        \State Draw a random sample $u_j \sim \text{Unif}(0,1)$
        \If{$u_j < p_j$}
            \State Keep the $j$-th observation in $\{\V y^{(k)}, \M X^{(k)}\}$.
        \Else
            \State Discard the $j$-th observation in $\{\V y^{(k)}, \M X^{(k)}\}$.
        \EndIf
    \EndFor
    
 \hskip -0.8 cm   \Return The selected source sample, denoted by $\{\V y^{(k-)}, \M X^{(k-)}\}$.
    
\end{algorithmic}
\end{algorithm}

\begin{figure}[t]
\captionsetup{font=footnotesize}
    \centering
     \includegraphics[width=\textwidth]{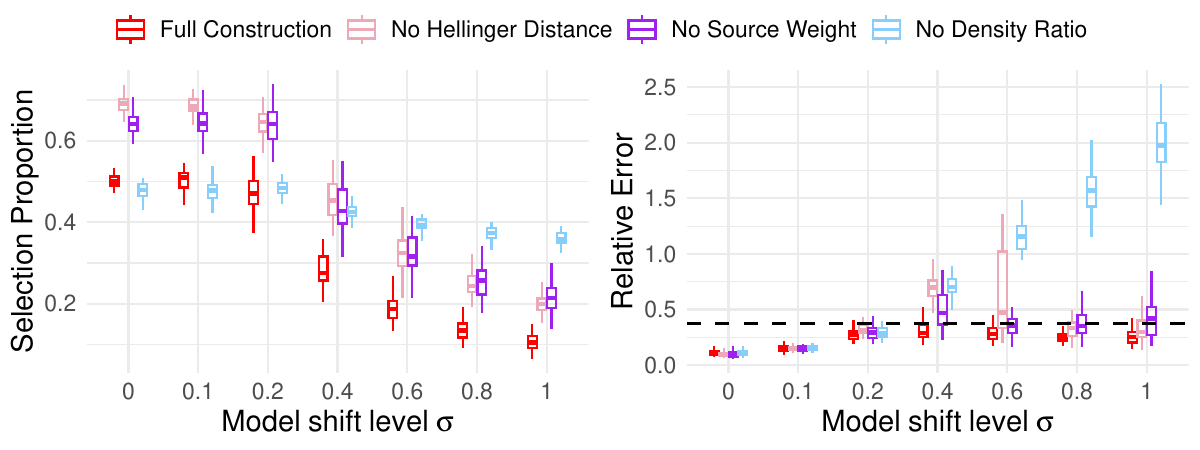}
    \caption{
    Boxplots of source selection proportions (left) and corresponding estimation errors (right) with varying model shift level $\sigma$ and 10\% outlier contamination for different constructions of $p_j$ over $50$ replicates. Outlier are present only in the source data.  Simulations follow the setup of Experiment 2 in Section \ref{sec:simulation_results} with a single source dataset of size $n = 1200$. The black dashed line (right panel) indicates the average estimation  error (0.38) of L\textsubscript{2}E fitted using only the target data.
    }
    \label{fig:selection_analysis}
\end{figure}

    To empirically justify the specific construction of the importance weight $p_j$ in \eqref{eq:sample-weight}, we conducted simulation experiments, following the setup of Experiment 2 in Section \ref{sec:simulation_results} with a single source dataset of size $n = 1200$, to evaluate source selection behavior under varying model shift ($\sigma$). We compared our full construction against three ablation variants that omit key components: the individual source weight $w_j^{(k)}$, the density ratio $\frac{\hat{f}_0(w^{(k)}_j)}{\hat{f}_k(w^{(k)}_j) + \epsilon}$,  and the cohort-level Hellinger distance term $\exp{(-\sqrt{h_k})}$. Figure \ref{fig:selection_analysis} presents the selection proportions and corresponding estimation errors (of TransL\textsubscript{2}E) under different model-shift levels and 10\% outlier contamination. As shown in the left panel of Figure \ref{fig:selection_analysis}, all formulations of $p_j$  exhibit a certain degree of under-selection, where our proposed full construction operates as a strategically conservative selection mechanism.
Under the scenario with no model shift ($\sigma=0$) and 10\% outlier contamination, the full construction retains about 50\% of the source samples. While ablation variants omitting the Hellinger distance or source weight are less conservative, selecting up to 70\% of the data, they offer no significant improvement in estimation error, as shown on the right panel. The advantage of the full construction becomes more pronounced as the model shift level increases. When $\sigma$ reaches 1, the selection proportion of the full construction drops sharply to about 15\%, whereas the ablation variants continue to select between 25\% and 40\% of the data. This over-inclusion leads to negative transfer, with estimation errors exceeding those of L\textsubscript{2}E fitted using only the target data. In contrast, the conservativeness of the full construction mitigates negative transfer, as evidenced by its consistently lower estimation errors compared to the target-only L\textsubscript{2}E.

Compared to hard thresholding methods, our proposed pseudo–importance-sampling approach in \Alg{TransL2E-selection} enjoys several merits. First, it selects the source data points that align well with the target samples based on the proposed importance weight \eqref{eq:sample-weight}, rather than relying on a threshold defined solely from the source data. 
Second, it selects data in a ``soft'' probabilistic manner. That is, source data points with high converged weights still have a non-zero probability of being excluded, while those with low weights retain a small chance of being selected. This probabilistic selection strategy accounts for between-cohort heterogeneity, particularly when $\V \beta^{(0)}$ and $\V \beta^{(k)}$ are not sufficiently close.
Finally, our proposed sampling procedure supports source selection at both individual and cohort levels.
In particular, if no observations are selected from a source, the resulting dataset $\{\V y^{(k-)}, \M X^{(k-)}\}$ is empty, thereby excluding the 
$k$-th source cohort and signaling it as non-transferable.


\textbf{Remark:} \textit{We note that \cite{zhao2023residual} and \cite{zheng2025transfer} have also explored individual-level data selection for transfer learning under linear regression and generalized linear models, respectively. Their approaches, however, differ substantially from ours and face several practical limitations. First, both methods assign an importance weight to each observation using density or probability ratio estimators, which require estimating the underlying density as well as obtaining an independent initial estimator of the regression coefficients. Second, sample selection in these methods is carried out using a hard threshold on the weights, rather than through a probabilistic mechanism as in our framework. The choice of threshold must be tuned via cross-fitting, and the required data splitting introduces additional challenges that can compromise the reliability and stability of the procedure.}


\section{Numerical Studies}
\label{sec:simulation}

We evaluated the empirical performance of the proposed TransL\textsubscript{2}E method over comprehensive simulation studies. 
We considered two different regression tasks: sparse linear regression and group sparse linear regression. In the main manuscript, we focused on sparse linear regression and demonstrated the performance of TransL\textsubscript{2}E by varying different problem parameters. For space consideration, simulation results for the group sparse regression setting were provided in the supplementary material. All simulation experiments were conducted in R. We implemented our TransL\textsubscript{2}E method in the \textsf{TransL2E} R package, which is  available at \hyperlink{https://github.com/Xiaoqian-Liu/TransL2E}{https://github.com/Xiaoqian-Liu/TransL2E}.


\subsection{Simulation setup}
\label{sec:sim-1}
\noindent\textbf{Data generation.} We simulated a target task as a sparse linear regression problem. Given the target sample size $n_0=200$ and the number of covariates $p=100$, we generated the $i$-th vector of covariates $\V x_i^{(0)}$, for $i=1, \cdots, n_0$, from standard multivariate normal distribution. The true coefficient vector $\V \beta_*^{(0)}$ was sparse, where its first $10$ elements were set to be $1$ and the remaining to be $0$.  The $i$-th response was then generated by $ y_i^{(0)} = {\V x_i^{(0)}}\Tra \V \beta_*^{(0)} + \epsilon_i$, where $\epsilon_i \sim N(0, {\tau^{(0)}}^{-2})$ was the Gaussian random noise. To simulate contamination, we shifted up the first $r_0n_0$ responses by $2\max(\V y^{(0)})$, where $r_0 \in [0, 1]$ was the contamination proportion of the target and $\max(\V y^{(0)})$ denoted the maximum value of the uncontaminated target responses. For the $K$ source datasets, each of them had a sample size of $n_k=400$ and shared the same number of covariates $p$ with the target. In each source $k$, we generated the $i$-th vector of covariates $\V x_i^{(k)}$ from the multivariate normal distribution $N(\V 0_p, \M \Sigma)$, where $\M \Sigma_{ij} = 0.5^{|i-j|}$.  The true coefficient vector $\V \beta_*^{(k)} = \V \beta_*^{(0)}+ \V \delta^{(k)}$, where $\delta_j^{(k)} \sim N(0, {\sigma^{(k)}}^2)$ for $j=1, \cdots, 30$ and $\delta_j^{(k)} =0$ otherwise. We referred to $\sigma^{(k)}$ as the \textit{model shift} level. The $i$-th response of the $k$-th source was  simulated by $ y_i^{(k)} = {\V x_i^{(k)}}\Tra \V \beta_*^{(k)} + \epsilon_i$ with $\epsilon_i \sim N(0, {\tau^{(k)}}^{-2})$, and the first $r_kn_k$ responses were lifted up by $2\max(\V y^{(k)})$ to introduce contamination, where $r_k \in [0,1]$ denotes the contamination proportion and $\max(\V y^{(k)})$ denoted the maximum value of the uncontaminated responses of the $k$-th source. We referred to $\nu_k = \frac{\tau^{(k)}}{\tau^{(0)}}$ as the \textit{precision shift} level.

\noindent\textbf{Competing methods.} We compared the proposed TransL\textsubscript{2}E method to the following five methods: (i) the Lasso regression \citep{tibshirani1997lasso} applied to the target dataset only,  referred to as Target-Lasso; (ii) the L\textsubscript{2}E regression \citep{L2E-Liu} with the Lasso penalty on the target dataset only, denoted by Target-L\textsubscript{2}E;  (iii) the L\textsubscript{2}E regression \citep{L2E-Liu} with the Lasso penalty applied to the pooled data (combining target and source datasets), termed Pooled-L\textsubscript{2}E; (iv)  the Trans-GLM method \citep{tian2023transfer} under the linear regression model with the Lasso penalty; and (v) the transfer learning quantile regression method with the Lasso penalty, named Trans-QR \citep{jin2024transfer}. While transfer learning for quantile regression is primarily designed to learn conditional quantiles, it also provides a robust estimation framework with the potential to handle contaminated data, and is therefore included in our comparison.
Note that among the three L\textsubscript{2}E methods, Target-L\textsubscript{2}E does not use any source data, Pooled-L\textsubscript{2}E incorporates the entire source data, and TransL\textsubscript{2}E, as a middle ground, selectively leverages high-quality source data identified by \Alg{TransL2E-selection}. 
Implementation details of the five competing methods are as follows. Target-Lasso was implemented using the \textsf{glmnet} R package \citep{glmnet}; Target-L\textsubscript{2}E and Pooled-L\textsubscript{2}E were implemented via the \textsf{L2E} R package \citep{L2E-Liu}; Trans-GLM was implemented using the \textsf{glmtrans} R package with its default source detection procedure \citep{tian2023transfer}, and Trans-QR was implemented through the \textsf{tfqr} R package with the informative set detection approach provided in \cite{jin2024transfer}.

\noindent\textbf{Tuning parameter selection.}
All methods in our simulation study have one or multiple tuning parameters $\lambda$ associated with the Lasso penalty. For Trans-GLM and Trans-QR, we used their default approaches to select the tuning parameters automatically. For the other four methods, we generated a decreasing sequence of $20$ $\lambda$ values
that were spaced evenly on a log scale from $10$ to $10^{-4}$ and then employed 5-fold cross-validation to select $\lambda$ from the sequence whenever such tuning parameter selection was needed. In addition, for Trans-QR, we fixed the quantile level at $0.5$ and did not treat it as a tuning parameter.

\noindent\textbf{Performance evaluation.}
We evaluated the competing methods on their performance in both coefficient estimation and variable selection. Specifically, to quantify the estimation accuracy, we used the relative  error, defined as $\text{relErr}=\lVert \hat{\V \beta}- \V \beta^T\lVert_2/\lVert \V \beta^T\lVert_2$, where $\hat{\V \beta}$ and $\V \beta^T$ are the estimated and true target coefficients, respectively. To measure the variable selection performance. we employed the F1 score \citep{L2E-Liu},  defined as $\frac{2\text{TP}}{2\text{TP}+\text{FP}+\text{FN}}$, where TP, FP and FN represent true positive, false positive, and false negative, respectively. The F1 score ranges from 0 to 1, with larger values indicating better variable selection performance.


To verify the practical efficacy of  TransL\textsubscript{2}E under different scenarios, we conducted various simulation experiments by varying different parameters. For each simulation
experiment, we ran 50 replicates for each method and used boxplots to report the median as well as the 25th and 75th quantiles of the two performance metrics.



\subsection{Simulation experiments and results}
\label{sec:simulation_results}
\noindent\textbf{Experiment 1: Varying outlier proportion.} As data contamination is the major focus of this work, we first examined whether TransL\textsubscript{2}E can handle outliers better than the other compared methods. To this end, we varied the source  contamination proportion $r_k$ among $\{0, 0.1, 0.2, 0.3, 0.4, 0.5\}$. For each setting of $r_k$, the specific parameters for data generation were as follows. For the target data, the precision parameter $\tau^{(0)}=1$ and the contamination proportion was fixed at $r_0=0.1$. The number of source datasets $K=5$, and for each source dataset, the model shift level $\sigma^{(k)}=0.2$ and the precision parameter $\tau^{(k)}=1$. 

\begin{figure}[t]
    \centering
    \captionsetup{font=footnotesize}
     \includegraphics[width=\textwidth]{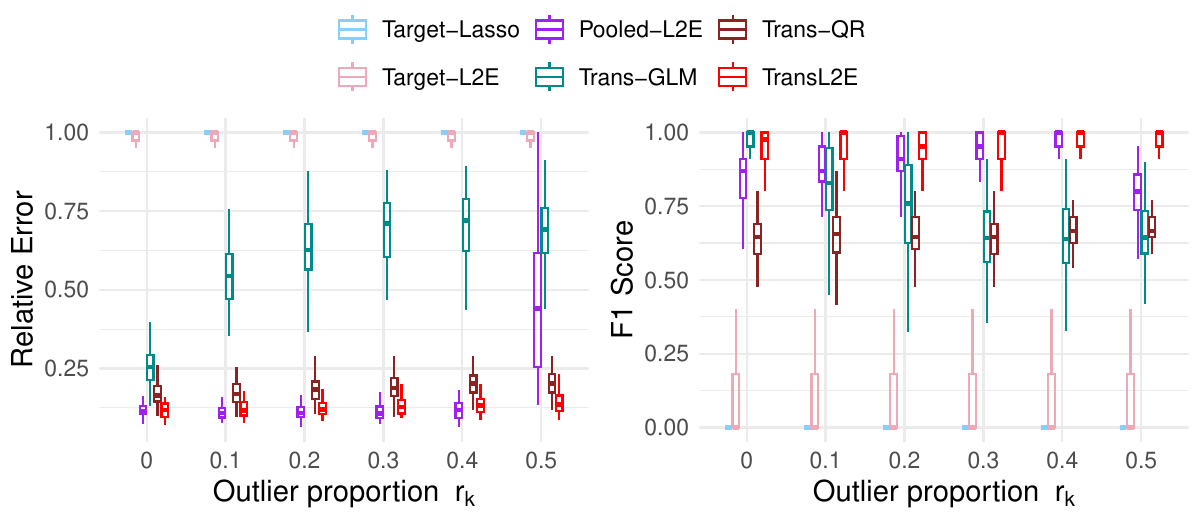}

    \caption{Results under varying outlier proportions in the source datasets.}
    \label{fig:out}
\end{figure}

\Fig{out} shows performances of all methods under different source outlier proportions. Only using the target dataset, both Target-Lasso and Target-L\textsubscript{2}E produced high estimation errors and low F1 scores, indicating that the target sample size was not sufficient for traditional models to produce accurate estimation. Target-L\textsubscript{2}E performed slightly better since it employed the L\textsubscript{2}E criterion to account for data contamination. With the source data available, Trans-GLM outperformed  Target-Lasso but its performance significantly degraded as the outlier proportion $r_k$ increased, reflected through its increasing relative errors and decreasing F1 scores. As a robust estimation method, Trans-QR achieved higher estimation accuracy than Trans-GLM but yielded lower F1 scores when the outlier proportion was low. We observed that TransL\textsubscript{2}E consistently exhibited superior performance in both coefficient estimation and variable selection across different outlier proportions, highlighting its ability in transferring useful information while accounting for data contamination. By leveraging the entire source data, Pool-L\textsubscript{2}E performed comparably to TransL\textsubscript{2}E when the outlier proportion in the source was low. However, when $r_k$ reached up to 0.5, Pool-L\textsubscript{2}E yielded substantially larger estimation errors and lower F1 scores, indicating the significant impact of the contaminated data. In contrast, through the bi-level  source detection procedure in \Alg{TransL2E-selection}, TransL\textsubscript{2}E identified high-quality samples to fit the model, resulting in robust performance against outliers.

\noindent\textbf{Experiment 2: Varying model shift level.} Model shift is a key concern for applying statistical transfer learning methods, since it cannot  be guaranteed in practice that both target and source share the same coefficients. To test how different methods perform under different model shift levels, we considered seven model shift settings where $\sigma^{(k)} = \{0, 0.1, 0.2, 0.4, 0.6, 0.8, 1\}$. Under each $\sigma^{(k)}$ setting, we set the target precision $\tau^{(0)}=1$, the target  contamination proportion $r_0=0.1$, and $K=5$ source datasets shared a common model shift $\sigma^{(k)}=\sigma$, precision parameter $\tau^{(k)}=1$, and contamination proportion $r_k=0.1$.   

\begin{figure}[t]
    \centering
    \captionsetup{font=footnotesize}
    \includegraphics[width=\textwidth]{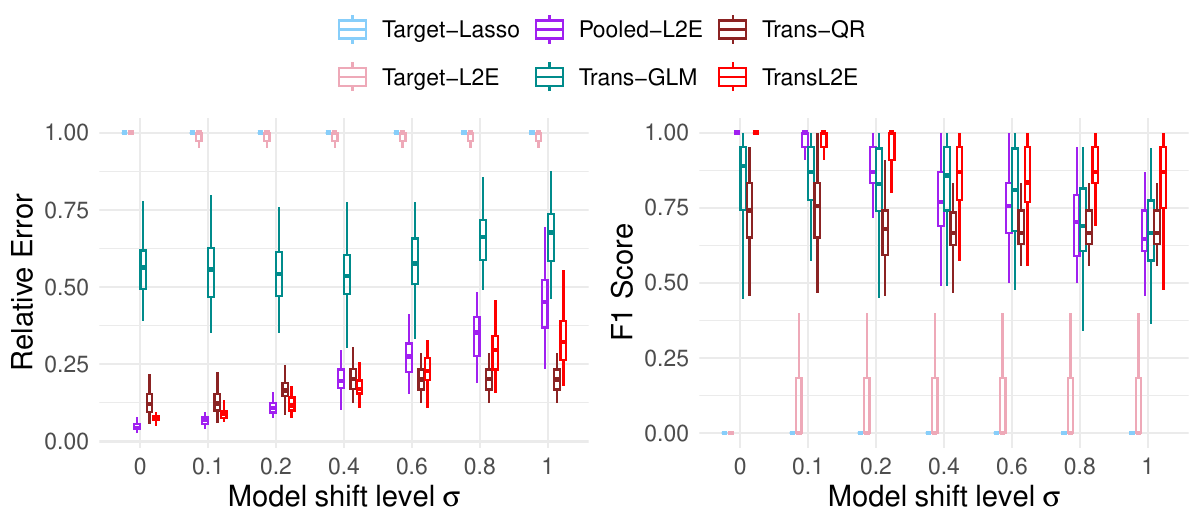}
    \caption{Results under varying model shift levels}
    \label{fig:ms}
\end{figure}

\Fig{ms} summarizes the results of all five  methods at varying model shift levels. Target-Lasso and Target-L\textsubscript{2}E exhibited poor performance due to the limited target sample size. With the additional source data, Trans-GLM achieved improved performance, but its estimation errors were high as it cannot handle outliers. For TransL\textsubscript{2}E and Pooled-L\textsubscript{2}E, their performance were comparable when the model shift level $\sigma$ was low. However, when $\sigma \geq 0.4$, Pooled-L\textsubscript{2}E yielded higher relative errors and lower F1 scores, reflecting the amplified influence by pooling all source datasets. In contrast, TransL\textsubscript{2}E's performance was less sensitive to model shift. It consistently produced lower estimation errors and higher F1 scores even when the model shift reached up to one, demonstrating its effectiveness in selectively transferring useful information. Trans-QR showed comparable or sometimes slightly higher estimation accuracy than TransL\textsubscript{2}E; however, its F1 scores were consistently lower, especially at low levels of model shift.

\begin{figure}[t]
    \centering
    \captionsetup{font=footnotesize}
    \includegraphics[width=\textwidth]{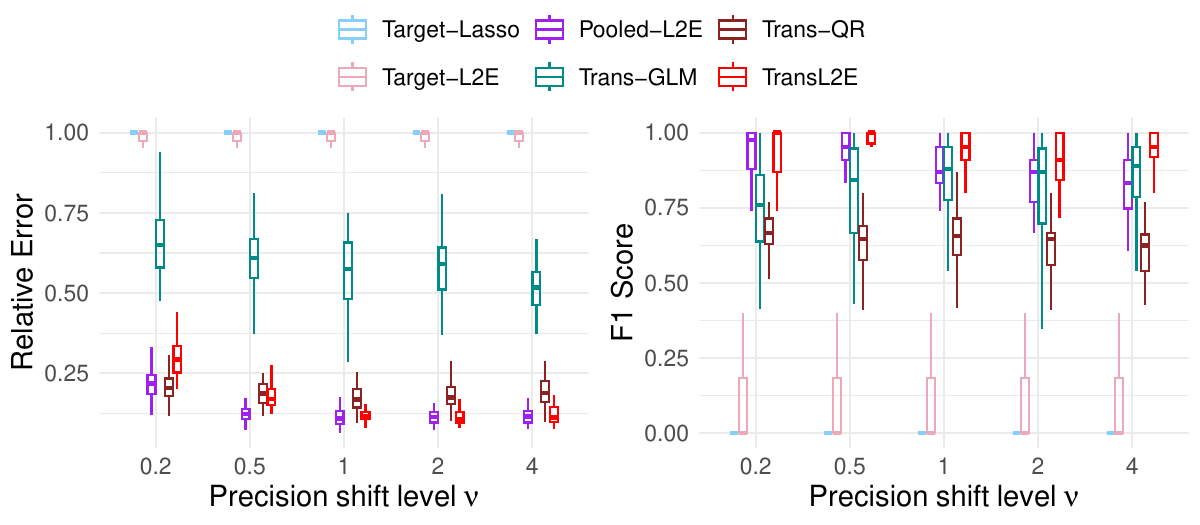}
    \caption{Results under varying precision shift levels}
    \label{fig:tau}
\end{figure}

\noindent\textbf{Experiment 3: Varying precision shift.} As discussed in Section \ref{sec:TransL2E}, the major focus of L\textsubscript{2}E regression is the coefficient parameter $\V \beta$, and we do not require that the target and source share  the same precision parameter. In fact, the precision parameter controls the signal-to-noise ratio (SNR) for the regression problem, and it is natural to allow the target and source problems to have different SNRs. To analyze how the precision parameter affects the method performance, we considered  varying the precision shift level over the grid $\{0.2, 0.5, 1, 2, 4\}$. Specifically, we generated a target dataset with  $\tau^{(0)}=1$ and  $r_0=0.1$, as well as $K=5$ source datasets with a shared  precision shift $\nu_k=\nu$ and $r_k=0.1$. The results under each value of $\nu$ were summarized in \Fig{tau}.

 As displayed in \Fig{tau}, the four methods (TransL\textsubscript{2}E, Pooled-L\textsubscript{2}E, Trans-GLM, and Trans-QR) utilizing the source data achieved improved estimation performance to varying extents as $\nu$ increased, since the SNR of the source data increases with $\nu$. TransL\textsubscript{2}E and Pooled-L\textsubscript{2}E produced lower relative errors and higher F1 scores compared to Trans-GLM, as they can handle data contamination through the L\textsubscript{2}E criterion.  While Trans-QR obtained estimation accuracy comparable to that of TransL\textsubscript{2}E and Pooled-L\textsubscript{2}E, it exhibited significantly lower F1 scores. Between the two L\textsubscript{2}E-based methods, TransL\textsubscript{2}E consistently outperformed Pooled-L\textsubscript{2}E in F1 scores, while achieving comparable estimation accuracy, especially when $\nu$ was large. 
 

\noindent\textbf{Experiment 4: Varying feature dimension.} 
In this experiment, we explore how the performance of different methods scale as the feature dimension $p$ varies. For that purpose, we considered three settings where $p=50, 250,$ and $500$, respectively. At each dimension setting, we generated the target data with sample size $n_0=200$, precision parameter $\tau^{(0)}=1$, and contamination proportion $r_0=0.1$. For $K=5$ source datasets, the data generation process remained the same as described in Section \ref{sec:sim-1} with the source precision $\tau^{(k)}=1$, model shift level  $\sigma^{(k)}=0.2$, and contamination proportion $r_k=0.1$ for $k=1, \cdots, K$. 

\begin{figure}[t]
    \centering
    \captionsetup{font=footnotesize}
    \includegraphics[width=\textwidth]{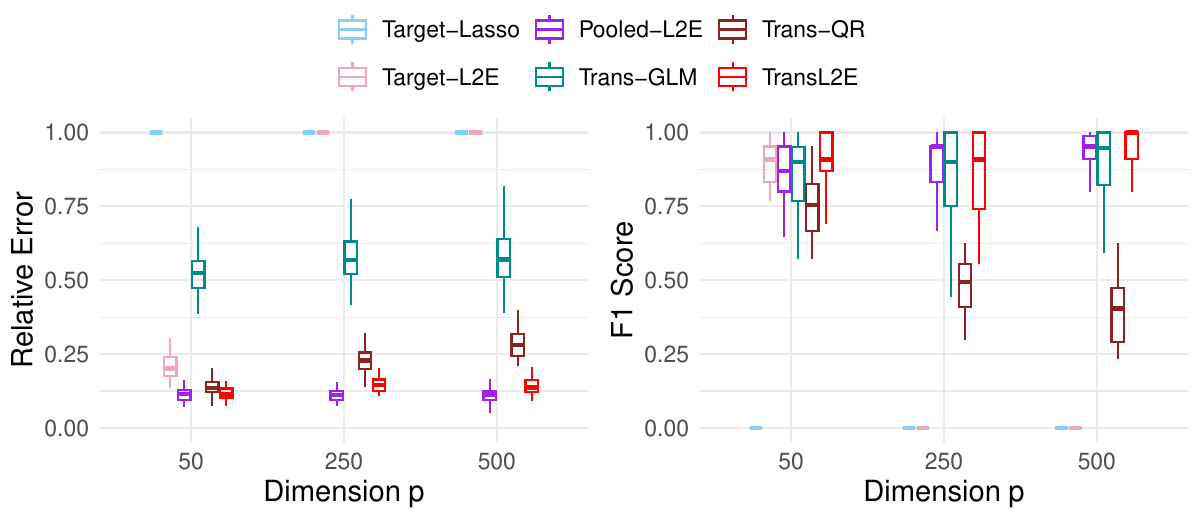}
    \caption{Results under varying feature dimensions}
    \label{fig:dim}
\end{figure}

\Fig{dim} displays the box plots of estimation errors and F1 scores of different methods at different feature dimensions. At the low-dimensional setting where $p=50$, Target-L\textsubscript{2}E produced estimation errors and F1 scores comparable with  Pooled-L\textsubscript{2}E and TransL\textsubscript{2}E, suggesting the target sample size was large enough to support target learning. For the moderate- ($p=250$) and high-dimensional ($p=500$) settings,  Target-L\textsubscript{2}E failed due to the limited sample size, whereas Pooled-L\textsubscript{2}E and TransL\textsubscript{2}E remained good performance by leveraging the source data. TransL\textsubscript{2}E obtained slightly higher F1 scores in the high-dimensional setting. 
Target-Lasso and Trans-GLM suffered from high estimation errors across different dimensional settings due to data contamination. Trans-QR attained estimation errors comparable to, or  higher than, those of Pooled-L\textsubscript{2}E and TransL\textsubscript{2}E, but exhibited substantially lower F1 scores in moderate- and high-dimensional settings.

\begin{figure}[t]
    \centering
    \captionsetup{font=footnotesize}
    \includegraphics[width=0.9\textwidth]{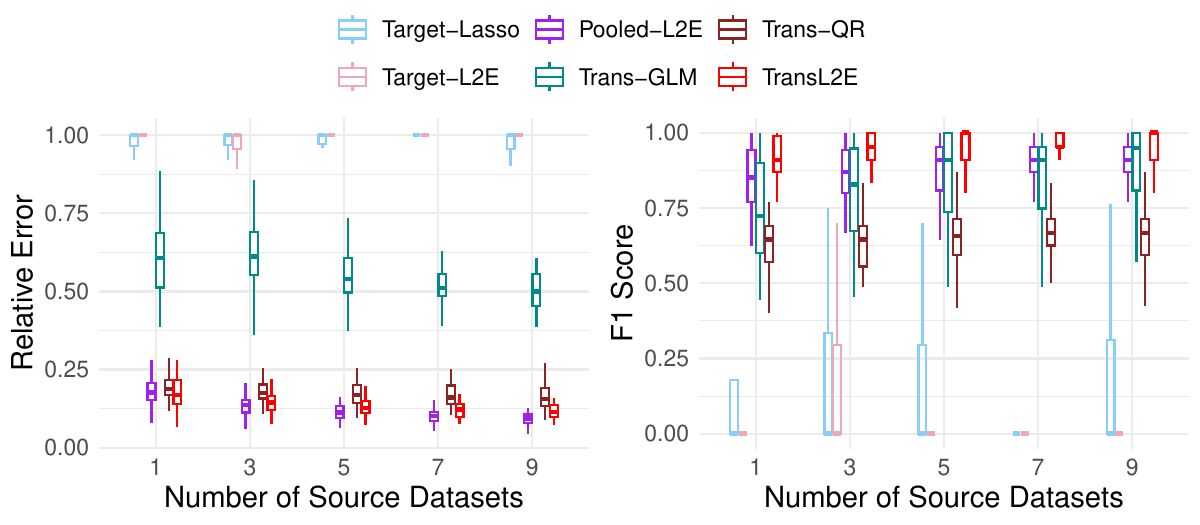}
    \caption{Results under varying numbers of source datasets.}
    \label{fig:source_number}
\end{figure}

\noindent\textbf{Experiment 5: Varying source number.}  We examined how different methods performed when more source data are available. Given a target dataset with the precision parameter $\tau^{(0)}=1$ and contamination proportion $r_0=0.1$. We generated $K$ source datasets with model shift level $\sigma^{(k)}=0.2$, precision parameter $\tau^{(k)}=1$, and contamination proportion $r_k=0.1$ for all $k=1, \cdots, K$. We set $K$ over the grid $\{1,3,5,7,9\}$ and summarized the results in \Fig{source_number}.

As expected, the four methods that utilized the source datasets achieved lower estimation errors and higher F1 scores as the number of source datasets increased. Trans-GLM yielded substantially higher errors than TransL\textsubscript{2}E and Pooled-L\textsubscript{2}E, since it was unable to handle outlier contamination. TransL\textsubscript{2}E performed comparably to Pooled-L\textsubscript{2}E in coefficient estimation but better in variable selection, as reflected by its higher F1 scores. Trans-QR's estimation performance was comparable to that of  TransL\textsubscript{2}E and Pooled-L\textsubscript{2}E, but its F1 scores were significantly lower.  Overall, TransL\textsubscript{2}E produced the best transfer learning performance by handling data contamination and source selection simultaneously.  

\section{Real Data Application}
\label{sec:real-data}
During the COVID-19 pandemic, county-level data were utilized to monitor the dynamics of disease spread and mortality \citep{zhang2020spatial}. Generally, mortality rates exhibit smooth spatial variation, as neighboring counties tend to share similar spatial and covariate effects. This spatial continuity facilitates the application of transfer learning approaches to borrow information from counties in the surrounding states, improving both the estimation of covariate effects and the prediction of mortality rates in the target state \citep{liuco}. However, this smooth pattern is not universal given that strict spatial homogeneity is rarely satisfied in practice. Some counties may exhibit mortality dynamics that deviate significantly from their neighbors due to localized outbreaks or recording discrepancies \citep{meirom2014localized, roukema2020anomalies}. These anomalies pose two challenges for effective transfer learning. On one hand, failing to account for these outliers leads to the incorporation of discordant information, inducing negative transfer and impacting prediction accuracy. On the other hand, simply excluding the whole data sources results in the loss of potentially useful samples, which could otherwise be utilized to improve target state estimation.

The proposed TransL\textsubscript{2}E method holds the potential to address the aforementioned challenges simultaneously. 
To validate its performance, we employed county-level COVID-19 data over 3,104 counties across the mainland U.S.\@. Mortality rates were aggregated and cleaned from various government and nonprofit institutions, and $37$ county-level covariates, including demographic, racial, socioeconomic, and comorbidity variables were selected for this study. Readers may refer to \citet{li2021identifying} for further details on data collection. For this analysis, we designated Florida (FL) as the target state and selected eight geographically surrounding states as the source states. Table~\ref{tab:target_source_summary} summarized the sample sizes for each state. Our objective was to (1) assess whether the proposed TransL\textsubscript{2}E method improves mortality rate prediction by leveraging auxiliary state data, and (2) examine how its source sample selection differs from that of conventional transfer learning methods, such as Trans-GLM and Trans-QR.

\begin{table}[t]
\captionsetup{font=footnotesize}
    \centering
    \begin{tabular*}{\textwidth}{@{\extracolsep{\fill}} l c c c c c c c c c }
        \toprule
        \textbf{State} & AL & AR & GA & LA & MS & NC & SC & TX & \textbf{FL} \\
        \midrule
        \textbf{Sample size} & 67 & 75 & 159 & 64 & 82 & 100 & 46 & 253 & 67 \\
        \bottomrule
    \end{tabular*}
    \caption{
    Target (bold) and source (regular) states in the COVID-19 study, with sample size representing the number of counties in each state.
    \label{tab:target_source_summary}}
\end{table}

\begin{table}[t]
\captionsetup{font=footnotesize}
    \centering
    \small
    \begin{tabular*}{\textwidth}{@{\extracolsep{\fill}} l c c c c c c c c c c}
        \toprule
        \textbf{Method} & Target-Lasso & Target-L\textsubscript{2}E & Pooled-L\textsubscript{2}E & Trans-GLM & Trans-QR & \textbf{TransL\textsubscript{2}E}  \\
        \midrule
        \textbf{Test MSE} & 0.59 & 0.22 & 0.17 & 0.14 & 0.15 & \textbf{0.12}  \\
        \bottomrule
    \end{tabular*}
    \caption{Comparison of MSE on the testing data for the mortality rate prediction of the target state (FL). \label{tab:test_mse_summary}}
\end{table}

As in our simulation study in Section~\ref{sec:simulation}, we compared the proposed TransL\textsubscript{2}E method against Trans-QR,  Trans-GLM, Pooled-L\textsubscript{2}E, and target-only baselines (Target-Lasso and Target-L\textsubscript{2}E). The target data were randomly split into 80\% training and 20\% testing sets. Table~\ref{tab:test_mse_summary} summarized the resulting Mean Squared Prediction Errors (MSE) on the target testing set. Notably, Target-L\textsubscript{2}E significantly outperformed Target-Lasso ($0.22$ vs. $0.59$), indicating the presence of data contamination in the target sample that necessitates robust estimation. While Trans-GLM and Trans-QR further reduced the error by leveraging source data, they remained vulnerable to negative transfer induced by outliers in the source states. The proposed TransL\textsubscript{2}E method achieved the lowest MSE, reinforcing its superior ability to transfer informative source knowledge while robustly mitigating the influence of outliers.

    
    
    

\begin{figure}[t]
\captionsetup{font=footnotesize}
    \centering
    \subfloat[\centering Trans-GLM Selection]{{\includegraphics[width=0.465\textwidth]{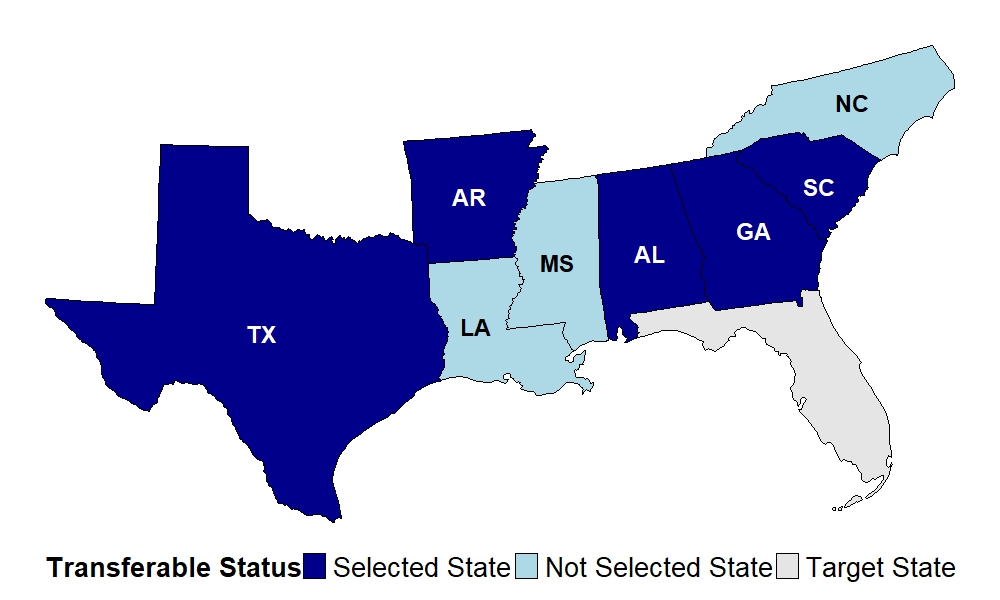} }}%
    \qquad
    \subfloat[\centering TransL\textsubscript{2}E Selection]{{\includegraphics[width=0.465\textwidth]{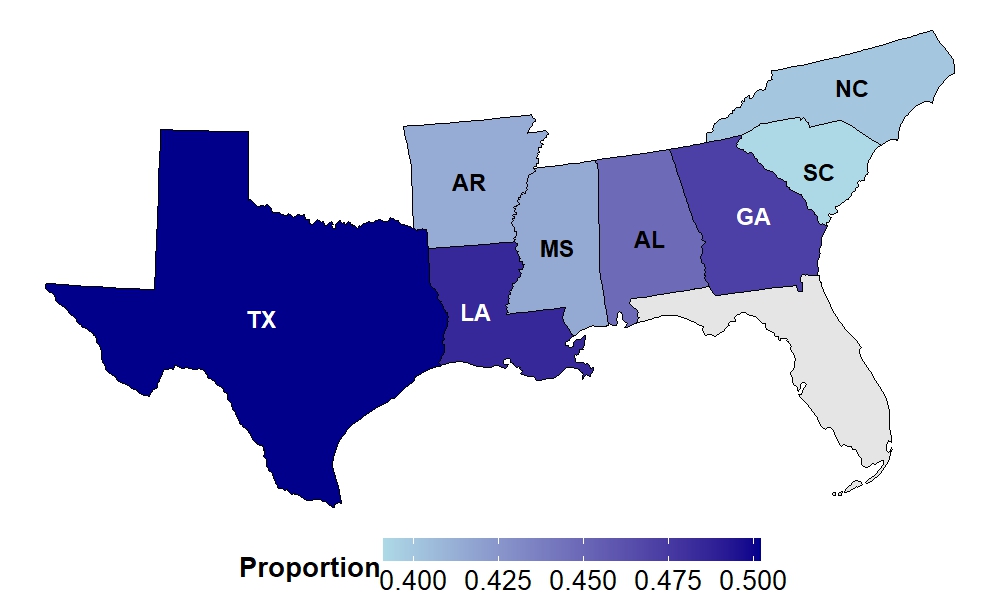} }}%
    \caption{Comparison of source data selection results. (a) Trans-GLM performed state-level binary selection. (b) TransL\textsubscript{2}E employed sample-level selection. The sample selection proportions of each source state for TransL\textsubscript{2}E are: AL: $0.45$, AR: $0.41$, GA: $0.47$, LA: $0.48$, TX: $0.50$, MS: $0.41$, SC: $0.39$, and NC: $0.40$.}%
    \label{fig:selection_maps}
\end{figure}

To better understand the performance disparities reported in Table~\ref{tab:test_mse_summary}, we analyzed the source data selection patterns across the three transfer learning methods. 
In this case, Trans-QR did not detect any source states as transferable and therefore reduced to a naive quantile regression fitted solely on target data, yielding the highest prediction MSE among the three methods. While Trans-GLM showed a slight performance improvement by incorporating source information, it retained only five of the eight candidate source states. Figure~\ref{fig:selection_maps} visualized the comparison of source data selection between Trans-GLM and TransL\textsubscript{2}E. As shown in panel (a), Trans-GLM employed a binary selection mechanism, which identified Louisiana (LA), Mississippi (MS), and North Carolina (NC) as ``non-transferable''  source states and excluded them completely.
In contrast, panel (b) revealed that the proposed TransL\textsubscript{2}E approach selected a significant portion of samples from  these regions (LA: $0.48$, MS: $0.41$, NC: $0.40$) that were identified as beneficial for the target estimation.
By retaining informative samples within globally “noisy” states, TransL\textsubscript{2}E prevented the loss of valuable training data that are overlooked by Trans-GLM, which potentially contributed to the improved test MSE reported in Table~\ref{tab:test_mse_summary}. These results further emphasized the effectiveness of the bi-level source detection procedure in identifying high-quality data points where the global selection methods fail to do so.

To further validate that the samples selected by TransL\textsubscript{2}E indeed contributed to the improved target task performance, we conducted a sensitivity analysis. Specifically, we evaluated the prediction MSE by applying the L\textsubscript{2}E method to the combined dataset of the target state (FL) and the specific subset of samples selected by TransL\textsubscript{2}E from each individual source state. As shown in Table~\ref{tab:l2e_source_MSE_summary}, incorporating the selected samples from any of the eight source states resulted in a lower MSE compared to the baseline L\textsubscript{2}E fitted on the target state alone (MSE: 0.22). Notably, including selected samples from GA or NC yielded a test MSE of 0.12, matching the aggregate performance of the full TransL\textsubscript{2}E framework that leverages all available source states. This suggests that the samples selected from these two states exhibit the strongest alignment with the target state (FL) and provide substantial predictive gains. More importantly, even for three source states that were entirely excluded by Trans-GLM as “non-transferable,” the inclusion of samples selected by TransL\textsubscript{2}E led to significant performance improvements over the baseline. These findings provide strong empirical evidence that the samples identified by our bi-level selection procedure yield positive transfer. By pinpointing high-quality data points within otherwise “noisy” source states, TransL\textsubscript{2}E effectively recovers valuable information that global selection methods fail to exploit, thereby explaining the superior overall performance observed in Table~\ref{tab:test_mse_summary}.

\begin{table}[t]
\captionsetup{font=footnotesize}
    \centering
    \begin{tabular*}{\textwidth}{@{\extracolsep{\fill}} l c c c c c c c c }
        \toprule
        \textbf{Source State} & AL & AR & GA & \textbf{LA} & \textbf{MS} & \textbf{NC} & SC & TX \\
        \midrule
        \textbf{L\textsubscript{2}E Test MSE} & 0.21 & 0.14 & 0.12 & \textbf{0.17} & \textbf{0.20} & \textbf{0.12} & 0.18 & 0.17 \\
        \bottomrule
    \end{tabular*}
    \caption{Test MSE for the target task using L\textsubscript{2}E fitted on the target state (FL) combined with the samples selected by TransL\textsubscript{2}E from each individual source state. The baseline L\textsubscript{2}E MSE using only target data is $0.22$. Bolded states highlight those are identified as negative transfer sources by Trans-GLM. 
    \label{tab:l2e_source_MSE_summary}}
\end{table}

\section{Discussion}
\label{sec:discussion}

In this article, we address data contamination in the context of statistical transfer learning. We propose TransL\textsubscript{2}E, a transfer learning framework for robust  structured regression. Built upon the L\textsubscript{2}E criterion, TransL\textsubscript{2}E is able to effectively handle outlier contamination in both target and source datasets. More importantly, it incorporates a novel data-driven bi-level source detection mechanism that selects high-quality source data, thereby ensuring reliable and beneficial information transfer. The proposed TransL\textsubscript{2}E framework is broadly applicable to various structured regression tasks. Through extensive simulation studies, TransL\textsubscript{2}E consistently outperforms existing methods across a wide range of scenarios.  A real data application to COVID-19 mortality prediction further highlights its practical utility in identifying informative source data at a fine-grained level while accounting for heterogeneity and contamination.

Several interesting directions for future research merit further investigation. First, a more thorough investigation of the source selection procedure is warranted. The current approach is relatively conservative, and an immediate question is whether more source data can be utilized without inducing negative transfer.
Second, the proposed bi-level source detection strategy has the potential to be extended to other transfer learning settings. 
Instead of fully including or entirely excluding a source dataset, selectively incorporating informative subsets of a source dataset may yield substantial benefits in many real-world applications. Lastly, a rigorous investigation of the statistical properties of the proposed TransL\textsubscript{2}E estimator would be of significant interest and would provide a valuable theoretical complement to the present work.

\section*{Acknowledgments}
Xiaoqian Liu was partially supported by the UCR  Academic Senate - Regents Faculty Fellowship (RFF) grant. Yang Feng was partially supported by the NSF Grant DMS-2324489. Haoming Shi was partially supported by the Ken Kennedy (2026) HPE Cray Graduate Fellowship.

\section*{Disclosure Statement}
\label{disclosure-statement}
The authors report there are no competing interests to declare.

\section*{Data Availability Statement}
\label{data-availability-statement}
The data that support the findings of this study are available in GitHub repository ``COVID-Health-Disparities'' at \url{https://github.com/lin-lab/COVID-Health-Disparities}.

\bibliographystyle{asa}

\bibliography{ref}







\end{document}